\documentclass[twocolumn,A4]{article}
\usepackage[dvips]{graphics}
\usepackage{setspace}
\usepackage{color}
\definecolor{gold}{rgb}{0.85,0.66,0}
\definecolor{dblue}{rgb}{0,0,0.8}
\begin{document}
\onecolumn
\begin{center}
{\bf{\Large {\textcolor{gold}{Electron transport through a quantum wire 
coupled with a mesoscopic ring}}}}\\
~\\
{\textcolor{dblue}{Santanu K. Maiti}}$^{1,2,*}$ \\
~\\
{\em $^1$Theoretical Condensed Matter Physics Division,
Saha Institute of Nuclear Physics, \\
1/AF, Bidhannagar, Kolkata-700 064, India \\
$^2$Department of Physics, Narasinha Dutt College,
129, Belilious Road, Howrah-711 101, India} \\
~\\
{\bf Abstract}
\end{center}
Electronic transport through a quantum wire sandwiched between two 
metallic electrodes and coupled to a quantum ring, threaded by a 
magnetic flux $\phi$, is studied. An analytic approach for the electron 
transport through the bridge system is presented based on the 
tight-binding model. The transport properties are discussed in three 
aspects: (a) presence of an external magnetic filed, (b) strength of 
the wire to electrode coupling, and (c) presence of in-plane electric 
field.

\vskip 1cm
\begin{flushleft}
\end{flushleft}
\vskip 4.7in
\noindent
{\bf ~$^*$Corresponding Author}: Santanu K. Maiti

Electronic mail:  santanu.maiti@saha.ac.in
\newpage
\twocolumn

\section{Introduction}
With the advancement in nanoscience and nanotechnology, the fabrication 
of sub-micron devices has become possible and has allowed one to study 
the electron transport through quantum systems in a very controllable 
way. These quantum systems have attracted much more attention since 
they constitute promising building blocks for future generation of 
electronic devices and directed attention on the study of discrete 
structures, such as a single molecule, arrays of molecules, quantum dots, 
quantum wires and mesoscopic rings. The electron transport through a bridge 
system was first studied theoretically in $1974$~\cite{aviram}. Later,
several numerous experiments~\cite{metz,fish,reed1,reed2,smit} have 
been performed through quantum systems placed between two metallic 
electrodes with few nanometer separation. The operation of such 
two-terminal devices is due to an applied bias. Current passing 
across the junction is strongly nonlinear function of applied bias 
voltage and its detailed description is a very complex problem. Though
lot of theoretical as well as experimental papers have been available
in the literature, yet the complete knowledge of the conduction mechanism 
in this scale is not well understood even today. The transport properties 
of these systems are associated with some quantum effects like, 
quantization of energy levels, quantum interference of electron waves, 
etc. A quantitative understanding of the physical mechanisms underlying 
the operation of nanoscale devices remains a major challenge in 
the present nanoelectronics research.

The aim of the present article is to reproduce an analytic approach 
based on the tight-binding model to investigate the electronic 
transport properties through a quantum wire coupled to a mesoscopic 
ring. There exist some {\em ab initio} methods for the calculation 
of conductance~\cite{yal,ven,xue,tay,der,dam}, yet it is needed the 
simple parametric approaches~\cite{muj1,muj2,sam,orella1,orella2,hjo,
baer1,baer2,baer3,walc1,walc2} for this calculation, especially for 
the case of larger molecular bridge systems. The parametric study is 
motivated by the fact that the {\em ab initio} theories are 
computationally too expensive and here we focus our attention on the 
qualitative effects rather than the quantitative ones. This is why we 
restrict our calculations on the simple analytical formulation of the 
transport problem.

We organize the paper as follow. Following the introduction 
(Section $1$), in Section $2$, we present the model system under 
consideration and give a very brief description for the calculation 
of conductance and current-voltage characteristics through the bridge 
system. Section $3$ presents the results of the system taken into 
account. Finally, we summarize our results in Section $4$.

\section{The model and a brief description onto the theoretical formulation}

\begin{figure}[ht]
{\centering \resizebox*{7.5cm}{3.75cm}{\includegraphics{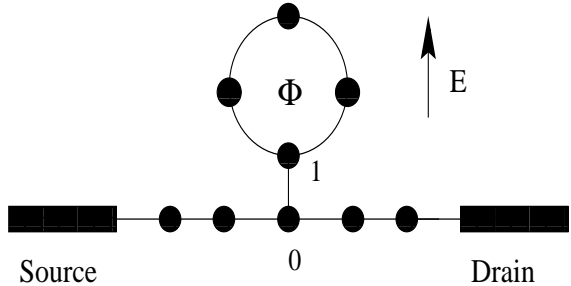}}\par}
\caption{Schematic view of a quantum wire coupled to a mescopic ring,
threaded by a magnetic flux $\phi$, and the wire is attached to two 
$1$D metallic electrodes.}
\label{molecule}
\end{figure}

We begin by referring to Fig.~\ref{molecule}. The system considered here 
is a quantum wire coupled to a mesoscopic ring with $N$ atomic sites and 
the wire is attached to two semi-infinite one-dimensional ($1$D) metallic 
electrodes, namely, source and drain.
The full system (quantum wire with ring) is described by a single-band 
tight-binding Hamiltonian within a non-interacting electron picture, and 
it can be written in the form,
\begin{equation}
H_C=H_W + H_R + H_{WR}
\label{equ1}
\end{equation}
where, $H_W$, $H_R$ and $H_{WR}$ correspond to the Hamiltonians for the
wire, ring and wire-to-ring coupling, respectively, and they can be
expressed as,
\begin{equation}
H_W=\sum_i \epsilon_i d_i^{\dagger} d_i + \sum_{<ij>}t_w 
\left(d_i^{\dagger}d_j + d_j^{\dagger}d_i \right)
\end{equation}
\begin{equation}
H_R=\sum_k \epsilon_k c_k^{\dagger} c_k + \sum_{<kl>}t_r 
\left(c_k^{\dagger}c_l e^{i\theta} + c_l^{\dagger}c_k e^{-i\theta} \right)
\end{equation}
\begin{equation}
H_{WR}=t_0 \left(c_1^{\dagger}d_0 + d_0^{\dagger}c_1 \right)
\end{equation}
Here, $\epsilon_i$'s ($\epsilon_k$'s) are the on-site energies of the 
ring (wire), $d_i^{\dagger}$ and $c_k^{\dagger}$ are the creation 
operators of an electron at site $i$ and $k$ in the wire and ring. 
$\theta=2 \pi \phi/N$ is the phase factor due to the flux $\phi$ 
threaded by the ring. $t_w$ ($t_r$) is the hopping integral between two 
nearest-neighbor sites in the ring (wire) and $t_0$ is the wire-to-ring 
tunneling coupling.

At much low temperatures and bias voltage, the linear conductance of 
the wire-ring system can be calculated by using one-channel Landauer 
conductance formula,
\begin{equation}
g=\frac{2 e^2}{h} T
\end{equation}
where $T$ is the transmission probability of an electron from the 
source to drain through the wire including the ring, and it is defined 
as~\cite{datta},
\begin{equation}
T(E,V)={\mbox{Tr}} \left[\left(\Sigma_S^r - \Sigma_S^a\right) 
G^r \left(\Sigma_D^a - \Sigma_D^r\right) G^a\right]
\label{trans1}
\end{equation}
Now the Green's function $G$ of the full system (wire with ring) is 
given by the relation,
\begin{equation}
G=\left[E-H_C-\Sigma_S-\Sigma_D\right]^{-1}
\label{equ7}
\end{equation}
where $E$ is the energy of injecting electrons from the source and $H$ is
the Hamiltonian of the full system described above (Eq.~\ref{equ1}). 
In Eq.~\ref{equ7}, $\Sigma_S=h_{SC}^{\dagger} g_S h_{SC}$ and 
$\Sigma_D=h_{DC} g_D h_{DC}^{\dagger}$, are the self-energy terms due to
the two electrodes. $g_S$ and $g_D$ correspond to the Green's functions 
for the source and drain, respectively. $h_{SC}$ and $h_{DC}$ are the 
coupling matrices and they are non-zero only for the adjacent points of 
the quantum wire and the electrodes. The coupling terms $\Gamma_S$ and 
$\Gamma_D$ for the full system can be calculated through the 
expression~\cite{datta},
\begin{equation}
\Gamma_{\{S,D\}}=i\left[\Sigma_{\{S,D\}}^r-\Sigma_{\{S,D\}}^a\right]
\end{equation}
where $\Sigma_{\{S,D\}}^r$ and $\Sigma_{\{S,D\}}^a$ are the retarded and
advanced self-energies respectively and they are conjugate to each
other. Datta {\em et al.}~\cite{tian} have shown that the self-energies
can be expressed like,
\begin{equation}
\Sigma_{\{S,D\}}^r=\Lambda_{\{S,D\}}-i \Delta_{\{S,D\}}
\label{equ9}
\end{equation}
where $\Lambda_{\{S,D\}}$ are the real parts of the self-energies which
correspond to the shift of the energy eigenvalues of the full system 
(quantum wire with ring) and the imaginary parts $\Delta_{\{S,D\}}$ of 
the self-energies represent the broadening of the energy levels. Since
this broadening is much larger than the thermal broadening, we restrict 
our all calculations only at absolute zero temperature. By doing some 
simple calculations, these real and imaginary parts of the self-energies 
can be determined in terms of the coupling strength ($\tau_{\{S,D\}}$) 
between the wire and two electrodes, injecting electron energy ($E$) 
and hopping strength ($v$) between nearest-neighbor sites in the 
electrodes. Using Eq.~\ref{equ9}, the coupling terms $\Gamma_S$ and 
$\Gamma_D$ can be written in terms of the retarded self-energy as,
\begin{equation}
\Gamma_{\{S,D\}}=-2 {\mbox{Im}}\left[\Sigma_{\{S,D\}}^r\right]
\end{equation}
All the information regarding the wire to electrode coupling are
included into the two self energies stated above and is analyzed 
through the use of Newns-Anderson chemisorption theory~\cite{muj1,muj2}. 
The detailed description of this theory is obtained in these two 
references.

Thus, by calculating the self-energies, the coupling terms $\Gamma_S$ and
$\Gamma_D$ can be easily obtained and then the transmission probability 
$T$ will be calculated from the expression given in Eq.~\ref{trans1}.

The current passing through the bridge is depicted as a single-electron 
scattering process between the two reservoirs of charge carriers. The 
current-voltage relation is evaluated from the following 
expression~\cite{datta},
\begin{equation}
I(V)=\frac{e}{\pi \hbar}\int \limits_{E_F-eV/2}^{E_F+eV/2} T(E,V)~ dE
\end{equation}
where $E_F$ is the equilibrium Fermi energy. For the sake of simplicity, 
here we assume that the entire voltage is dropped across the 
wire-electrode interfaces and this assumption does not greatly affect 
the qualitative aspects of the $I$-$V$ characteristics. Throughout the 
article we set $E_F$ to $0$ and use the units $c=e=h=1$.

\section{Results and discussion}

Here we describe conductance-energy and current-voltage characteristics 
through the quantum wire coupled to a mesoscopic ring at absolute zero 
temperature. Electron transport properties through the system are strongly 
affected by the magnetic flux $\phi$, wire-to-electrode coupling strength 
and the in-plane electric field. In the presence of in-plane electric filed 
and assuming it along the perpendicular direction of the wire, the dependence 
of the site energies on the electric field $\mathcal E$ is written within 
the tight-binding approximation as~\cite{orella1},
\begin{eqnarray}
\epsilon_i &=& \left(e\mathcal{E} a N/2 \pi \right) \cos\left[2\pi(i-1)/N
\right] \nonumber \\
 & = & \left(e t_r \right) \left(\mathcal{E}^{\star} N/2 \pi \right)
\cos\left[2\pi(i-1)/N\right]
\end{eqnarray}
where, $a$ is lattice spacing in the mesoscopic ring and 
$\mathcal{E}^{\star}$ is the dimensionless electric field strength 
defined by $\mathcal{E}a/t_r$. For simplicity, here we assume $t_w$, $t_r$ 
and $t_0$ are identical to each other in magnitude and specify them by the 
symbol $t$. We investigate all the essential features of electron transport 
for the two limiting cases. One is the weak-coupling limit, defined as 
$\tau_{\{S,D\}} << t$ and the other one is the 
\begin{figure}[ht]
{\centering \resizebox*{8cm}{8cm}{\includegraphics{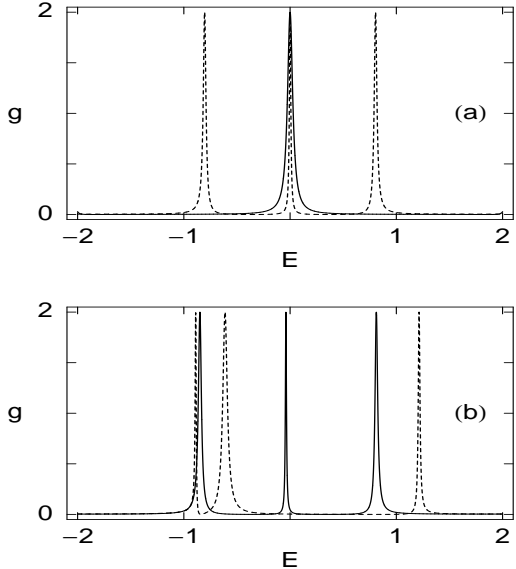}}\par}
\caption{Conductance $g$ as a function of energy $E$ in the weak-coupling 
limit for the system with ring size $N=10$, where (a) in the absence of 
any electric filed with $\phi=0$ (solid line) and $0.4$ (dotted line) and 
(b) in the presence of $\phi=0.4$ with $\mathcal E=2$ (solid line) and 
$4$ (dotted line).}
\label{elecondlow}
\end{figure}
strong-coupling limit and defined it as $\tau_{\{S,D\}} \sim t$. The 
parameters $\tau_S$ and $\tau_D$ correspond to the couplings of the 
wire to the source and drain, respectively. The common set of values 
of these parameters in the two limiting cases are as follow: 
$\tau_S=\tau_D=0.5$, $t=3$ (weak-coupling) and $\tau_S=\tau_D=2$, $t=3$ 
(strong-coupling).

In Fig.~\ref{elecondlow}, we plot the conductance ($g$) as a function of 
the injecting electron energy ($E$) for the bridge system in the limit of
weak-coupling. Figure~\ref{elecondlow}(a) corresponds to the spectrum in 
the absence of any electric filed where, the solid and dotted curves are 
respectively for $\phi=0$ and $0.4$. In Fig.~\ref{elecondlow}(b), the 
spectrum is shown for the non-zero value of the electric field with 
$\phi=0.4$ where, the solid and dotted curves represent the results for 
the electric filed strengths $\mathcal E=2$ and $4$, respectively. 
Conductance vanishes almost for all energies except at resonances where 
it approaches to $2$. At these resonances, the transmission probability 
$T$ becomes unity, since $g=2T$ (from the Landauer formula with $e=h=1$).
The resonant peaks in the conductance spectrum coincide with eigenenergies 
of the system (wire including the ring), and thus the spectrum manifests 
itself the energy levels of the system. For zero electric field strength 
and in the absence of magnetic flux $\phi$, the conductance exhibits 
a single resonant peak across $E=0$ (see solid curve of 
Fig.~\ref{elecondlow}(a)), while, in the presence of $\phi$ more resonant 
peaks appear in the spectrum (see dotted curve of Fig.~\ref{elecondlow}(a)). 
It reveals that for non-zero value of $\phi$ more resonating states 
appear in the system. This is due to the removal of all the degeneracies 
in the energy eigenstates for any non-zero value of $\phi$. In the 
presence of in-plane electric field, these resonant peaks are shifted 
and the conductance spectrum becomes asymmetric with respect to the 
energy $E$ (see Fig.~\ref{elecondlow}(b)).

\begin{figure}[ht]
{\centering \resizebox*{8cm}{8cm}{\includegraphics{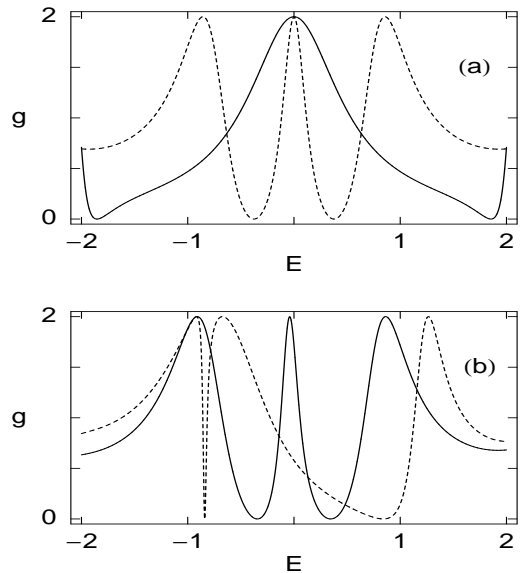}}\par}
\caption{Conductance $g$ as a function of energy $E$ in the strong-coupling 
limit for the system with ring size $N=10$, where (a) in the absence of 
any electric filed with $\phi=0$ (solid line) and $0.4$ (dotted line) and 
(b) in the presence of $\phi=0.4$ with $\mathcal E=2$ (solid line) and 
$4$ (dotted line).}
\label{elecondhigh}
\end{figure}

For the strong wire-to-electrode coupling, resonant peaks get substantial 
widths as presented in Fig.~\ref{elecondhigh} where, the solid and dotted 
curves
correspond to the identical meaning as earlier. The increment of the 
resonant widths is due to the broadening of the energy levels of the
wire including the ring, where the contribution comes from the imaginary 
parts of the two self-energies~\cite{datta}.

The scenario of electron transfer through the bridge becomes much more 
clearly visible by studying the current $I$ as a function of the 
applied bias voltage $V$.
\begin{figure}[ht]
{\centering \resizebox*{7.5cm}{4.5cm}{\includegraphics{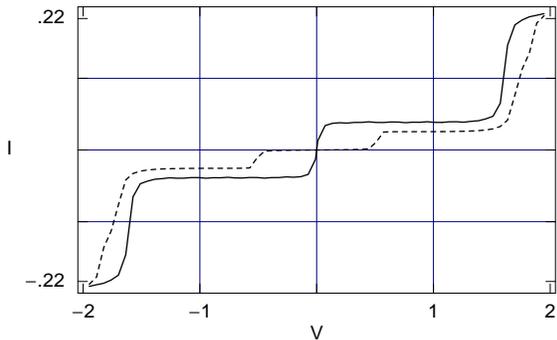}}\par}
\caption{Current $I$ as a function of bias voltage $V$ in the limit of 
weak wire-to-electrode coupling for the system with ring size $N=10$ 
and $\phi=0.4$. The solid and dotted lines correspond to the currents 
for $\mathcal E=0$ and $3$, respectively.}
\label{elecurrlow}
\end{figure}
The Current is computed from the integration procedure of the transmission
function $T$ which shows the same variation, differ only in magnitude by 
the factor $2$,
\begin{figure}[ht]
{\centering \resizebox*{7.5cm}{4.5cm}{\includegraphics{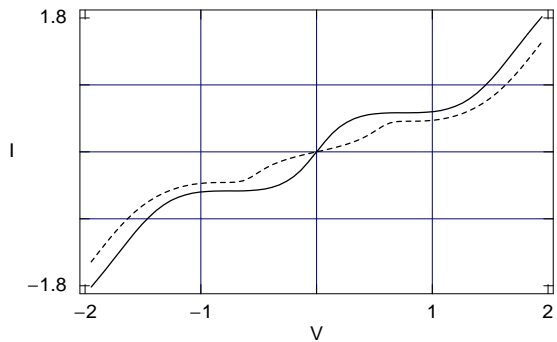}}\par}
\caption{Current $I$ as a function of bias voltage $V$ in the limit of 
strong wire-to-electrode coupling for the system with ring size $N=10$ 
and $\phi=0.4$. The solid and dotted curves correspond to the currents for 
$\mathcal E=0$ and $3$, respectively.}
\label{elecurrhigh}
\end{figure}
like as the conductance spectra (Figs.~\ref{elecondlow} and
\ref{elecondhigh}). The current-voltage characteristic in the 
weak-coupling limit for the bridge system is shown in Fig.~\ref{elecurrlow}
where, the solid curve corresponds to the current in the absence of any 
electric field and the dotted curve denotes the same for $\mathcal E=3$. 
Here we take $\phi=0.4$. The current shows staircase-like behavior with 
sharp steps, which is associated with the discrete nature of the resonant 
spectrum (Fig.~\ref{elecondlow}). The shape and width of the current steps 
depend on the width of the resonant spectrum since the hight of a step in 
$I$-$V$ curve is directly proportional to the area of the corresponding 
peak in the conductance spectrum. On the other hand, the current varies 
continuously with the applied bias voltage and achieves much bigger values 
in the strong-coupling limit, as shown in Fig.~\ref{elecurrhigh} where, 
the solid and dotted curves correspond to the same meaning as earlier. 
From both Figs.~\ref{elecurrlow} and \ref{elecurrhigh} it is clearly 
observed that the in-plane electric field suppresses the current 
amplitude (see the dotted curves). This feature may be utilized to 
control {\em externally} the amplitude of the current through the 
bridge system.

\section{Concluding remarks}

To summarize, we have introduced parametric approach based on the 
tight-binding model to investigate the electron transport properties 
at absolute zero temperature through a quantum wire coupled to a 
mesoscopic ring threaded by a magnetic flux $\phi$. A simple parametric 
approach is given to study electron transport properties through the
system, and it can be used to study the transport behavior in any 
complicated molecular bridge system. Electronic conduction through 
the quantum wire is strongly influenced by the flux $\phi$ threaded
by the ring and the wire-to-electrode coupling strength. The effects 
of in-plane electric field have also been studied in this context and 
it has been predicted that the current amplitude can be controlled 
{\em externally} through the bridge system by means of this electric 
field.

\end{document}